\documentclass{emulateapj}

\shorttitle{The All-Sky X-ray Selected Cluster Dipole}
\shortauthors{Kocevski \& Ebeling}

\begin{document}

\title{On The Origin of the Local Group's Peculiar Velocity}
\author{Dale D. Kocevski and Harald Ebeling}

\affil{Institute for Astronomy, University of Hawaii, 2680 Woodlawn Dr., Honolulu, HI 96822}
\email{kocevski@ifa.hawaii.edu; ebeling@ifa.hawaii.edu}

\begin{abstract}
 
We aim to settle the debate regarding the fraction of the Local Group's peculiar velocity that is induced by structures beyond the Great Attractor by calculating the dipole anisotropy of the largest, all-sky, truly X-ray selected cluster sample compiled to date.  The sample is the combination of the REFLEX catalog in the southern hemisphere, the eBCS sample in the north, and the CIZA survey in the Galactic plane.  The composite REFLEX+eBCS+CIZA sample overcomes many of the problems inherent to previous galaxy and cluster catalogs which limited their effectiveness in determining the origin of the Local Group's motion.  From the dipole anisotropy present in the cluster distribution we determine that $44\%$ of the Local Group's peculiar velocity is due to infall into the Great Attractor region, while $56\%$ is in the form of a large-scale flow induced by more distant overdensities between $130$ and $180$ $h^{-1}$ Mpc away.  In agreement with previous analyses, we find that the Shapley supercluster is the single overdensity most responsible for the increase in the dipole amplitude beyond $130$ $h^{-1}$ Mpc, generating $30.4\%$ of the large-scale contribution.  Despite the dynamical significance of both the Great Attractor and Shapley regions, we find that additional superclusters play an important role in shaping the Local Group's peculiar velocity.  Locally, the Perseus-Pisces region counteracts much of the Great Attractor's effect on the acceleration field beyond $60$ $h^{-1}$ Mpc.  At larger distances we find that numerous groupings and loose associations of clusters at roughly the same distance as the Shapley region induce a significant acceleration on the Local Group. These include the well known Horologium-Reticulum concentration, as well as newly noted associations centered on  Abell 3667 and Abell 3391 and a string of CIZA clusters near C1410 which may trace an extension of the Shapley complex into the Zone of Avoidance.  We also note the presence of a significant underdensity of clusters in the northern hemisphere roughly $150$ $h^{-1}$ Mpc away and suggest that the large-scale anisotropy observed in the cluster distribution near this distance may have as much to do with the presence of large overdensities in the south as it does with the lack of superclusters in the north.  Finally we discuss reasons for the discordant results obtained using cluster and galaxy samples in determining the origin of the Local Group's motion.
\end{abstract}


\section{Introduction}

It is generally accepted that the dipole signature in the cosmic microwave background (CMB) is due to a Doppler effect arising from the motion of the Local Group (LG) through the cosmological reference frame.  Although the direction and amplitude of this motion is known to high accuracy, its source has yet to be conclusively determined.   While linear perturbation theory predicts that the LG's peculiar velocity is induced by anisotropies in the surrounding matter distribution, there has been disagreement regarding the distance out to which inhomogeneities in the density field continue to affect the LG's dynamics.  The debate has centered on whether the LG is principally accelerated by a massive, nearby attractor which has remained hidden behind the Galactic plane, or whether a significant portion of its motion is in the form of a large-scale bulk flow induced by more distant structures, as has been suggested by an increasing number of studies.  Resolving the source of the LG's motion and the bulk flow in which it participates carries many interesting cosmographical and cosmological implications.  For example, in order for a distant supercluster like the Shapley concentration (Shapley 1930) to induce an infall at the LG as large as the one produced by the more nearby Hydra-Centaurus complex, its mass would need to be sixteen times greater than that of the largest overdensity observed in the local volume.  Furthermore, if distant structures contribute to the LG's peculiar velocity, then anisotropies in the large-scale matter distribution must exist to at least those structures, implying that the universe becomes isotropic only at larger distances.  

\subsection{The Great Attractor}

Early attempts to determine the source of the LG's peculiar velocity largely suggested a local origin.   The nearest large-scale overdensity, the Virgo cluster, which generates a 240 km s$^{-1}$ infall velocity at the LG (Jerjen \& Tammann 1993), accounts for 27\% of the LG's 627 km s$^{-1}$ velocity toward the CMB dipole (Kogut 1993).  Virgo's inability to explain all of the LG's motion led Shaya (1984) to suggest an additional flow toward the Hydra-Centaurus supercluster.  This flow was subsequently detected by Lynden-Bell et al. (1988, hereafter LB88) as a systematic distortion in the peculiar velocities of 400 early-type galaxies within $40 h^{-1}$ Mpc and its amplitude at the LG was estimated to be 570 km s$^{-1}$.  Assuming this motion was due to infall into a single ``Great Attractor'' (GA), LB88 estimated that the source of the flow was located roughly $43 h^{-1}$ Mpc away, or about $13 h^{-1}$ Mpc behind Centaurus.  This distance, coupled with the large infall velocity, implied the rather high mass of $\sim5\times10^{16} h^{-1}_{50}$ M$_{\odot}$ for the GA complex.  The LB88 findings suggested that the GA was responsible for the remaining $\sim70\%$ of the LG's peculiar velocity not induced by the Virgo cluster. 

Despite the LB88 findings, subsequent redshift surveys that have encompassed the GA region have failed to detect a mass overdensity as large as the one implied by the LB88 peculiar velocity data (Dressler 1988; Strauss et al. 1992; Hudson 1993, 1994), nor have they conclusively measured the backside infall into the GA one would expect if the region were best described as a single, stationary attractor (Mathewson et al. 1992, Courteau et al. 1993).  Even with the more recent discoveries of rich clusters such as Norma (Abell 3627, Kraan-Korteweg et al. 1996) and CIZA J1324.7-5736 (Ebeling, Mullis \& Tully 2002) near the Hydra-Centaurus region, there remains a significant discrepancy between the mass concentration observed in the GA and the mass originally proposed by LB88 (Staveley-Smith et al. 2000, Kocevski et al. 2006).  

\subsection{The Large-Scale Contribution}

There is now a growing volume of work suggesting the dynamical significance of the GA was originally overestimated and that some component of the LG's peculiar velocity is in the form of a large-scale bulk flow which continues past the GA region and is induced by more distant structures.  One of the first suggestions of a non-local component was put forth by Plionis (1988), who proposed that the anisotropy observed in the galaxy number-counts of the Lick catalog (Shane \& Wirtanen 1967) implied a portion of the LG's motion originating from beyond $\sim80 h^{-1}$ Mpc.  More recently, Hudson et al. (2003) show that results from the peculiar velocity studies of Hudson et al. (1999), Dale et al. (1999), Willick (1999), and Colless et al. (2001) are all consistent with a 350 km s$^{-1}$ bulk flow continuing beyond $60 h^{-1}$ Mpc, ruling out nearby overdensities such as the GA as the source of the motion.  Likewise, Zaroubi et al. (1999) show that if the local velocity field is decomposed into its divergent (locally produced) and tidal (externally produced) components, only $50\%$ of the flow toward the GA is due to infall into the Hydra-Centaurus region, while the remaining velocity is due to a continuing bulk flow generated by attractors beyond $80 h^{-1}$ Mpc.  In addition, Tonry et al. (2000), using surface brightness fluctuation (SBF) distances, find a mass for the GA that is $\sim6$ times less than the original LB88 estimate.  Their best-fit models suggest the GA is well centered on the Centaurus cluster (as opposed to $13 h^{-1}$ Mpc behind it) and that part of the flow toward the GA is due to a $\sim150$ km s$^{-1}$ residual bulk motion.  Tonry et al. essentially propose that the LG's motion, which was mistakenly identified as infall into a single, massive attractor slightly behind Centaurus, is in fact the result of two flows, one into the Hydra-Centaurus region and a second flow toward a more distant source.  

Many of the studies finding evidence for a continuing flow beyond the GA have suggested that some fraction of the motion may be due to infall into the Shapley supercluster (SSC), located $\sim100 h^{-1}$ Mpc behind the Hydra-Centaurus complex.  The SSC region is unique in the local universe, containing the richest concentration of clusters out of all the 220 identified superclusters out to $z = 0.12$ (Einasto et al. 1997).  In fact, the SSC contains more than 4 times the number of rich clusters present in the GA region.  The SSC's possible dynamical significance was pointed out by Scaramella et al. (1989, 1991) and Plionis \& Valdarnini (1991) because of the high concentration of Abell clusters in the region and its directional alignment with the GA.  Since the two regions are only separated by $\sim24^{\circ}$ on the sky, Kocevski et al. (2004) suggest this alignment causes a bootstrap effect that sets in place the large-scale density anisotropy that is responsible for the LG's peculiar velocity.  

Despite the significant overdensity present in the SSC, its dynamical impact on the LG has been a matter of debate.  Rowan-Robinson et al. (2000) use the dipole anisotropy present in the IRAS Point Source Catalog Redshift sample (PSCz, Saunders et al. 1995) to conclude that the SSC has a marginal influence on the LG, inducing only $\sim20$ km s$^{-1}$ of its motion.   Likewise, Erdogdu et al. (2005), using the dipole anisotropy of the 2 Micron All-Sky Redshift Survey (2MRS, Huchra et al. 2005), find that structures beyond $140 h^{-1}$ Mpc induce only a negligible acceleration on the LG.  On the other hand, Lucey et al. (2005) combine the cluster peculiar velocity data from the SMAC (Hudson et al. 1999), SCI/SCII (Giovanelli et al. 1999), ENEARc (Bernardi et al. 2002) and SBF (Tonry et al. 2000) surveys and find that both the GA and the SSC generate an equal amount of the LG's peculiar velocity.  This higher estimate concurs with the results from dipole analyses of various cluster samples; Branchini \& Plionis (1996) and Plionis \& Kolokotronis (1998) use the dipole anisotropy in the distribution of optically selected Abell/ACO (Abell 1958, Abell et al. 1989) and X-ray Brightest Abell-type Cluster (XBAC, Ebeling et al. 1996) samples, respectively, to determine that $\sim32\%$ of the LG's motion is due to the SSC.  Kocevski et al. (2004) recently added X-ray selected clusters from the CIZA survey (named for Clusters In the Zone of Avoidance, Ebeling, Mullis, \& Tully 2002) to the XBAC distribution to fill in the Zone of Avoidance (ZOA) and found an even larger SSC contribution ($\sim50\%$).   

\subsection{The All-Sky X-ray Selected Cluster Dipole}

In this study we aim to settle the debate regarding the fraction of the LG's peculiar velocity that is induced by structures beyond the GA by calculating the dipole anisotropy of the largest, all-sky, truly X-ray selected cluster sample compiled to date.  The sample is the combination of the ROSAT-ESO Flux Limited X-ray catalog (REFLEX, B\"{o}hringer et al. 2004) in the southern hemisphere, the extended Brightest Cluster Sample (eBCS, Ebeling et al. 1998, 2000) in the north, and the CIZA survey in the Galactic plane.  The composite REFLEX+eBCS+CIZA sample (hereafter RBC) overcomes many of the problems inherent to previous galaxy and cluster catalogs which limited their effectiveness in determining the origin of the LG's motion.  First of all, due to its X-ray selected nature, the RBC sample maps the distribution of massive X-ray luminous clusters which tend to be the accelerators of large-scale flows.  In this sense, the RBC sample traces the peaks of the density fluctuation field better than galaxy samples such as the PSCz, which has been shown to undersample dense regions such as the SSC (Kaiser et al. 1991).  X-ray bright clusters are also luminous enough for statistically complete samples to be constructed out to larger distances than galaxy samples, which face an increasing incompleteness beyond 60 $h^{-1}$ Mpc.  We suspect that the estimate for a marginal SSC acceleration from the PSCz data may be a result of these two factors.

Second, the X-ray selected RBC sample overcomes several limitations of early cluster catalogs such as Abell/ACO, which are optically selected.  Since optical selection methods rely on identifying individual clusters through overdensities in the projected galaxy distribution and assigning cluster masses based on the size, or richness, of those overdensities, fluctuations in the surface density of field galaxies as well as superpositions of poor clusters or filamentary structure along the line of sight can lead to false detections or overestimates of a system's richness (e.g. van Haarlem 1997, see Sutherland 1988).  These projection effects work to systematically amplify the measured dipole amplitude.  In addition, optical cluster searches suffer from severe extinction and stellar obscuration in the direction of the Milky Way, leading to catalogs with poor coverage in a $40^{\circ}$ wide strip centered on the plane of the Galaxy.  This is particularly troubling since large-scale structures associated with the GA and the SSC are known to exist in, or extend into, the ZOA (Tully et al. 1992).   A variety of techniques have been used to reconstruct the ZOA, ranging from uniform filling (Strauss \& Davis 1988; Lahav 1987) to a spherical-harmonics approach which extends structures above and below the plane into the ZOA (Plionis \& Valdarnini 1991, cf. Brunozzi et al. 1995), but the value of these reconstruction techniques is limited if the Milky Way does indeed obscure dynamically significant regions, as has been suggested.  The X-ray selected nature of the RBC sample makes it preferable over its optically selected counterparts since (1) cluster X-ray emission originates from the $\sim10^{7}$ K intracluster medium, which is more peaked at the gravitational center of the cluster than the projected galaxy distribution. This minimizes projection effects since clusters would need to be in almost perfect alignment to be mistaken for a single, more luminous object, (2) X-ray luminosity is closely correlated with cluster mass (Reiprich \& B\"{o}hringer 1999), thus providing a better estimate to a system's dynamical impact than a clusters' projected galaxy richness, and (3) X-ray emission does not suffer as severe an extinction in the plane of the Galaxy (Ebeling, Mullis \& Tully 2002). 

Finally, the RBC's relatively low flux limit makes it a better tracer of large-scale structures than previous X-ray confirmed cluster samples, such as the XBAC catalog which has recently been used by Plionis \& Kolokotronis (1998) and Kocevski et al. (2004) to investigate the origin of the LG's motion.  Due to the XBAC's fairly high X-ray flux limit, the sample is limited to only the most massive clusters, which, although tracing the deepest potential wells, only sparsely sample the underlying density field and can lead to an increased level of shot-noise.  In addition, although X-ray confirmation effectively eliminates projection effects in XBAC, the catalog remains optically selected, therefore clusters missed by the Abell/ACO sample will not be included in the XBAC sample.  This essentially means very nearby, very extended clusters are systematically missed as they often do not contrast strongly with the background galaxy population.  The undersampling of clusters at low redshifts would lead to an overestimate of the contribution to the dipole from distances greater than 60 $h^{-1}$ Mpc, where the XBAC incompleteness is minimal.  The RBC's X-ray flux limit is nearly half of the one used in the XBAC survey, which means the sample is no longer limited to extremely massive clusters.  This increased depth more than triples the number of clusters present in the RBC sample, which in turn affords us a greater resolution in tracing the overdensities which give rise to the LG's motion.  

The RBC sample is currently the largest, most complete X-ray selected cluster sample for which to trace the large-scale structure of the local universe and determine the origin of the LG's peculiar velocity.  In what follows we use the RBC sample to (1) determine the role of the cluster distribution in producing the LG's peculiar velocity, (2) determine the fraction of the LG's motion that is induced from distances greater than 60 $h^{-1}$ Mpc, and (3) estimate the relative contribution of various superclusters such as the GA and SSC regions, to the final dipole amplitude.  We proceed in the following manner: in \S2 we give an overview of the REFLEX, eBCS, and CIZA samples, \S3 describes the details of the dipole analysis, and our results are put forward in \S4.  Finally we summarize our primary conclusions in section 5.  Throughout this paper we assume an Einstein-de Sitter universe with $q_{0}=0.5$ and $H_{0}=100$ $h$ km s$^{-1}$Mpc$^{-1}$ unless otherwise stated, so that our results are directly comparable to those of previous studies.

\section{Data}

We aim to measure the dipole anisotropy present in the all-sky, X-ray selected cluster sample created by combining the REFLEX, eBCS and CIZA catalogs.  In this section we describe the attributes of the REFLEX, eBCS and CIZA samples in further detail and the methods used to homogenize them into one all-sky catalog.

\subsection{The REFLEX Sample}

The REFLEX catalog is the most comprehensive X-ray selected cluster sample compiled for the southern hemisphere, consisting of 447 clusters with X-ray fluxes greater than $3\times 10^{-12}$ erg cm$^{-2}$ s$^{-1}$ in the 0.1--2.4 keV band.  The survey is limited to declinations of $\delta < 2.5^{\circ}$, redshifts of $z\leq 0.3$ and Galactic latitudes away from the Galactic plane ($|b|>20^{\circ}$).  REFLEX hails from the ROSAT All-Sky Survey (RASS, Voges 1992) and is therefore truly X-ray selected.  As described in B\"{o}hringer et al. (2001), the RASS Bright Source Catalog (Voges et al. 1999) provided 54076 X-ray detections which were reexamined by means of a growth curve analysis (B\"{o}hringer et al. 2000), resulting in 1417 targets with X-ray fluxes above $3\times 10^{-12}$ erg cm$^{-2}$ s$^{-1}$.  Cross-correlating these targets with galaxy overdensities in the COSMOS optical database (MacGillivray \& Stobie 1984) produced 673 candidate clusters.  Subsequent screening and follow-up observations led to 447 confirmed galaxy clusters in the final REFLEX sample.  Estimated to be over 90$\%$ complete, REFLEX is the largest cluster catalog produced from the RASS to date.

\subsection{The eBCS Sample}

The eBCS catalog is the most complete X-ray flux-limited cluster sample constructed from the RASS for the northern hemisphere, consisting of 290 clusters with X-ray fluxes greater than $3\times 10^{-12}$ erg cm$^{-2}$ s$^{-1}$ in the 0.1--2.4 keV band.  The sample is limited to declinations of $\delta > 0^{\circ}$ and redshifts of $z\leq 0.3$ and, like REFLEX, the survey avoids the Galactic plane ($|b|>20^{\circ}$).  The eBCS sample was compiled by cross-correlating the RASS with existing cluster catalogs such as the Abell/ACO and Zwicky (Zwicky et al. 1961-68) samples, while also using the Voronoi Tessellation and Percolation (VTP, Ebeling 1993; Ebeling \& Wiedenmann 1993) algorithm to allow for an improved determination of basic cluster characteristics, as well as to detect additional clusters based on their X-ray properties alone.  The fact that the eBCS is not selected from the RASS Bright Source Catalog in the same manner as REFLEX leads to slight differences in the global properties of the two samples; corrections for this and other systematic effects are discussed in \S2.4.  The eBCS catalog is estimated to be 75\% complete within the 2190 deg$^{2}$ processed by the VTP algorithm.

\subsection{The CIZA Sample}

The CIZA sample is the product of the first systematic search for X-ray luminous clusters behind the plane of the Galaxy.  As described in Ebeling, Mullis \& Tully (2002, hereafter EMT), CIZA targets were selected from the RASS Bright Source Catalog if they met three criteria: (1) location in the ZOA, $| b | < 20^{\circ}$, (2) an X-ray flux greater than $1\times 10^{-12}$ erg cm$^{-2}$ s$^{-1}$ (0.1--2.4 keV) and (3) a spectral hardness ratio exceeding a preset threshold value\footnote{The minimum hardness ratio threshold depends on location in the plane; see EMT for details.} to discriminate against softer, non-cluster sources.  The resulting target list of 1901 sources was cross-correlated with existing databases to identify known clusters and remove obvious non-clusters.  The remaining cluster candidates were then subjected to a comprehensive imaging and spectroscopic follow-up campaign.  The use of the Bright Source Catalog as a target list means CIZA is not correlated with any optically selected catalog and is therefore truly X-ray selected.  

A subsample of 73 CIZA clusters with fluxes above $5\times 10^{-12}$ erg cm$^{-2}$ s$^{-1}$ (the B1 sample) has recently been published (EMT).  To allow the combination of CIZA with REFLEX and eBCS, we use the total fluxes of the B1 sample listed by EMT and also add clusters from a second, yet unpublished extended sample, which includes all clusters whose total fluxes within a metric 1.5 Mpc aperture exceed $3\times 10^{-12}$ erg cm$^{-2}$ s$^{-1}$.  The resulting sample, limited to $z<0.3$ and $|b| < 20^{\circ}$, contains 151 clusters.  

\begin{figure}[t]
\epsscale{1.2}
\plotone{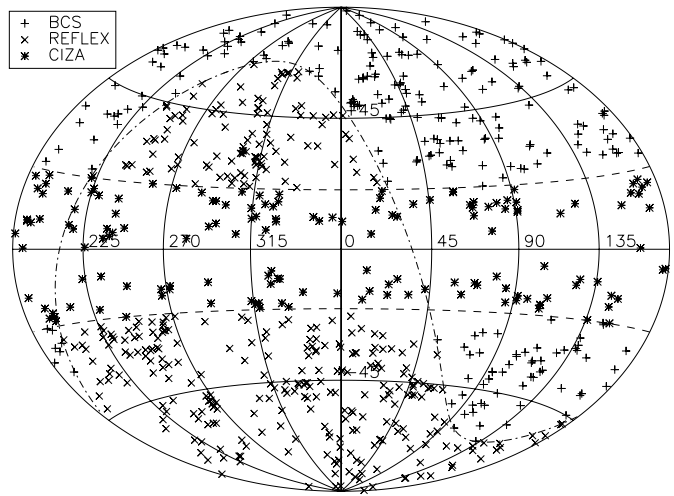}
\caption{Aitoff projection of the combined REFLEX+eBCS+CIZA cluster sample in Galactic Coordinates.  The dashed lines represent the traditional ZOA ($|b|<20^{\circ}$), while the dashed-dotted line is the celestial equator ($\delta=0^{\circ}$).}
\end{figure}

\subsection{Sample Homogenization}

\begin{figure}[t]
\epsscale{1.2}
\plotone{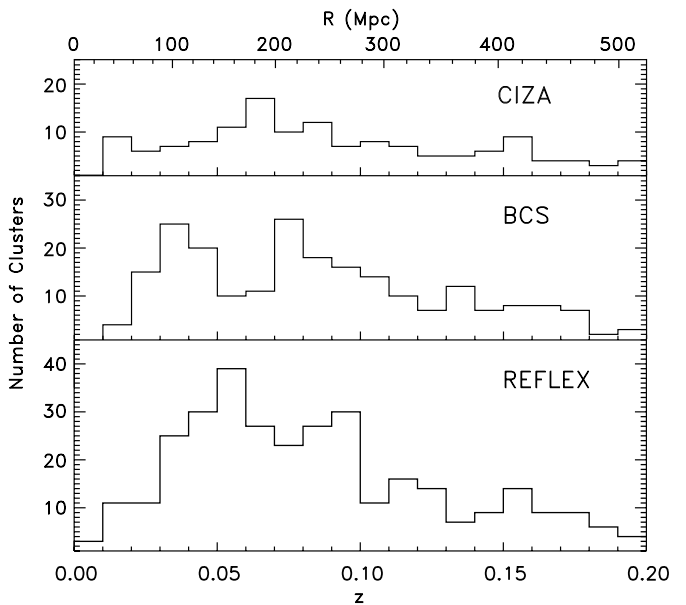}
\caption{The redshift distribution of the 359 REFLEX, 248 eBCS, and 151 CIZA clusters with recomputed X-ray fluxes above $f_{x} \geq 3\times 10^{-12}$ erg cm$^{-2}$ s$^{-1}$ in the 0.1--2.4 keV band}
\end{figure}

\begin{figure}[t]
\epsscale{1.2}
\plotone{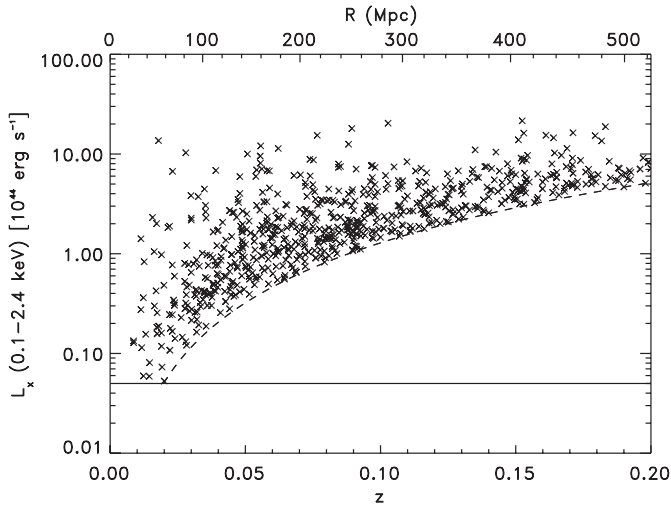}
\caption{The redshift-luminosity distribution of the 758 RBC clusters.  The dashed line denotes the sample's X-ray flux limit of $3\times 10^{-12}$ erg cm$^{-2}$ s$^{-1}$ (0.1--2.4 keV) while the solid horizontal line denotes the sample's lower luminosity limit of $5\times 10^{42}$ $h^{-2}$ ergs s$^{-1}$.  The sample is volume complete out to 59 Mpc, where the dashed and solid lines intersect.}
\end{figure}

The combined effects of differing selection techniques, flux measuring algorithms and other systematic effects make merging the REFLEX, eBCS and CIZA catalogs a non-trivial task.  The method used to combine the individual catalogs into one homogeneous all-sky sample will be described in greater detail in a forthcoming paper and is therefore only briefly discussed here.

First we need to ensure that the X-ray flux of each cluster is measured in a consistent manner.  The need for uniformly measured fluxes is two-fold: it ensures that (1) all three samples are complete to the same depth and (2) the weight given to each cluster in the dipole analysis, which is derived from the cluster's X-ray luminosity, is determined in the same manner throughout the sky.  The depth and weighting of the sample must be considered carefully when measuring the dipole anisotropy, since systematic differences in either can introduce spurious contributions to the dipole amplitude and pointing.  As published, REFLEX fluxes are determined using a growth curve analysis which corrects for any X-ray flux missed outside the detection aperture by extrapolating out to the cluster's estimated virial radius.  Conversely, the eBCS flux correction extrapolates out to infinity, which leads to an $8.3\%$ difference in the fluxes measured by the two approaches (B\"{o}hringer et al. 2004).  Finally, CIZA fluxes are measured by summing all of the X-ray emission within a metric 1.5 Mpc radius (at the cluster redshift) about the centroid of the cluster's X-ray emission.  In this study, we adopt this latter approach and recalculate the flux of all REFLEX, eBCS and CIZA clusters using a metric 1.5 Mpc aperture.  Using RASS data, we redetermine the centroid of each cluster's X-ray emission, remove point sources within the detection aperture, and calculate a new X-ray count rate at the location of each cluster, taking into account the local RASS exposure time.  The X-ray background at each position is determined in a 1 Mpc wide annulus (2-3 Mpc from the cluster centroid) about the measurement aperture and subtracted from the observed count rate.  The measured count rates are then converted to unabsorbed fluxes in the 0.1-2.4 keV band by taking into account the line-of-sight interstellar hydrogen column density as given by Dickey \& Lockman (1990).  Next, clusters whose X-ray emission appears to be dominated by a point source are removed if more than 50\% of their X-ray emission originates from within the central 10\% of the detection aperture.  Finally a flux cut at $3\times 10^{-12}$ erg cm$^{-2}$ s$^{-1}$ is applied, leaving 359 REFLEX, 248 eBCS, and 151 CIZA clusters in the resulting sample\footnote{This total includes a pair of double clusters counted as two single objects}.  The distribution of the combined sample in Galactic coordinates is shown in Figure 1 and its redshift distribution is shown in Figure 2.  The distribution of cluster luminosities versus redshift is shown in figure 3.

\begin{figure}[t]
\epsscale{1.2}
\plotone{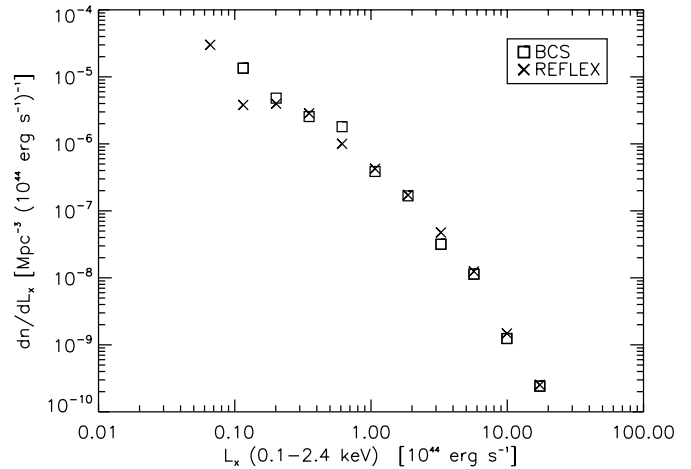}
\caption{The binned REFLEX and eBCS X-ray luminosity functions.  The luminosity functions are computed using our recalculated X-ray fluxes and corrective weights for the eBCS.}
\vspace*{1mm}
\end{figure}

In addition to the different flux measuring procedures of the three catalogs, differences also exist in the selection technique employed.  Whereas REFLEX and CIZA had the Bright Source Catalog source detection algorithm run on RASS data covering their entire survey area, the eBCS employed the VTP to search for clusters only over regions which were correlated with previously known clusters.  While this led to the detection of both known and unknown clusters, the serendipitously detected clusters could only be found in the 2190 deg$^{2}$ which were processed with VTP.  Given the frequency of serendipitously detected clusters in the surveyed area (17 in 2190 deg$^{2}$), 84 additional clusters should exist with fluxes above $3\times 10^{-12}$ erg cm$^{-2}$ s$^{-1}$ in regions of the northern sky not processed by the eBCS.  Since REFLEX, by design, should detect all such serendipitous clusters in its survey area, a density variation is introduced between the northern and southern portions of the combined sample.  Left uncorrected such a systematic density difference will bias the dipole pointing toward the southern portion of the sample.  To compensate for this effect in the dipole analysis, we weight each eBCS cluster by the difference in the comoving cluster density between the eBCS and REFLEX.  This is essentially equivalent to weighting each eBCS cluster by a factor of $w_{x} = 1.34$, which is the multiplicative correction needed to make up for the 84 clusters not included in the eBCS catalog.  This weighting scheme is similar to the one employed to correct the density variation between the Abell and ACO cluster samples (Brunozzi et al. 1995, Plionis \& Kolokotronis 1998), although our weighting is distributed uniformly throughout the eBCS since we do not expect the missing clusters to be correlated with any particular part of the sky or with redshift.  To check the effectiveness of this weighting scheme, we fit a Schechter function to the X-ray luminosity functions of the REFLEX and eBCS portions of our sample and find that the best-fit parameters agree quite well between the two sections.   The binned luminosity functions, which were constructed using our recalculated X-ray fluxes and the corrective weighting for the eBCS, are shown in Figure 4.  

Finally, systematic effects in the CIZA sample are corrected for in much the same manner as for the eBCS.  The catalog's primary incompleteness comes from the fact that clusters are systematically missed in the very central regions of the ZOA, which is due to the increased difficulty of obtaining spectroscopic confirmation of clusters through the severe extinction within $\pm5^{\circ}$ of the Galactic plane and toward the Galactic center.  To correct for this we again weight each CIZA cluster by the difference in the comoving cluster density between the CIZA and REFLEX samples, which amounts to a factor of $w_{x} = 1.63$.  Given the relatively sparse nature of cluster samples and the known clustering of clusters, we find that distributing the weight over the entire sample is preferable to adding random clusters to the Galactic plane.  To ensure that the weighting of eBCS and CIZA clusters in this manner does not overly influence our findings, we calculate the dipole with our weights set to unity and find that our results are quite robust to variations in the adopted weighting scheme.

\section{Methodology}

Linear theory of gravitational instability dictates that the peculiar velocity of a reference frame can be related to the gravitational acceleration induced by the mass distribution surrounding it via

\begin{equation}
\textit{\textbf{v}}_{p} \hspace{.1in} = \hspace{.1in} \frac{H_{o}\beta}{4\pi \bar{n}} \int \frac{n(r)}{r^{2}} \textit{\textbf{\^{r}}}\ dr 
\end{equation}

\noindent (Peebles 1976), where $\beta = \Omega^{0.6}_{0}/b$ and $b$ is the biasing parameter relating the mass tracers to the underlying mass distribution they represent, and $\bar{n}$ is the average mass-tracer number density.  In other words, Equation 1 tells us that the dipole moment of a mass-tracer distribution can be directly related to the peculiar velocity that sample would induce on the LG.   For a non-continuous, flux-limited tracer sample, such as a cluster catalog, a more useful version of equation (1) is

\begin{equation}
\textit{\textbf{v}}_{p} = \frac{H_{o}\beta}{4\pi \bar{n}} \hspace{.05in} \sum_{i=1}^{N} \frac{w_{i}}{\phi(r_{i})r_{i}^{2}}\textit{\textbf{\^{r}}}_{i}
\end{equation}
\begin{eqnarray}
\hspace{.55in} = \hspace{.1in} \beta \textit{\textbf{D}}_{\rm cl} \hspace{0.28in} \nonumber
\end{eqnarray}

\noindent where $r_{i}$ is the distance to each cluster, $\phi(r_{i})$ is the sample's selection function at  $r_{i}$, $w_{i}$ is weight assigned to the $i$th cluster, and $\textbf{\emph{\^{r}}}_{i}$  are the unit vectors pointing to the position of each cluster.  $\textit{\textbf{D}}_{\rm cl}$ is the vector quantity we refer to as the dipole throughout the rest of this paper.  From equation (2) it can be seen that the characteristics of the dipole are such that its amplitude will increase with distance until the largest inhomogeneity in the sample is encompassed and isotropy is reached, after which the dipole flattens out to its final value. This flattening signals that the convergence depth, $R_{\rm conv}$, has been reached.  Assuming that $\textit{\textbf{D}}_{\rm cl}$ is well aligned with the LG's peculiar velocity, $\textit{\textbf{v}}_{p}$, at this convergence depth, equation (2) provides the means to estimate the $\beta$ parameter.  The rest of this section is devoted to describing the distance, selection function and cluster weights used in equation (2).

To convert our observed redshifts to distances, we use the formula of Mattig (1958):
\begin{equation}
        \nonumber    r = \frac{c}{H_{0}q_{0}^{2}(1+z)}[q_{0}z+(1-q_{0})(1-\sqrt{2q_{0}z+1})],
\end{equation}
which reduces to 
\begin{equation}
	             r = \frac{2c}{H_{0}}\left(1-\frac{1}{\sqrt{z+1}}\right)
\end{equation}
for our assumed value of $q_{0} = 0.5$.  To account for possible peculiar-velocity contamination in our redshifts, which may alter the perceived distance to a cluster, we follow the approach of Kocevski et al. (2004) and perform our analysis in both the LG and CMB reference frames, since it has been shown that the cluster dipole calculated in these frames over- and under-estimate, respectively, the real-space dipole (Branchini \& Plionis 1996).   We transform our measured heliocentric redshifts, $z_{\odot}$, into the LG and CMB rest frames using 
\begin{equation}
                            cz_{_{\rm LG}} = cz_{\odot}+300\sin l \sin b
\end{equation}
and
\begin{eqnarray}
  cz_{_{\rm CMB}} = cz_{_{\rm LG}}+v_{_{\rm LG}}\hspace{0.02in}\big[\hspace{0.02in}\sin(b)\sin(b_{_{\rm CMB}})\\ +\hspace{0.05in} \cos(b)\cos(b_{_{\rm CMB}})\cos(|l_{_{\rm CMB}}-l|)\hspace{0.01in}\big] \nonumber
\end{eqnarray}
where $v_{_{\rm LG}}$ is the amplitude of the LG velocity as inferred from the CMB dipole anisotropy and $(l_{_{\rm CMB}},b_{_{\rm CMB}})$ is the direction of this motion in Galactic coordinates.  Authors have traditionally removed the LG's Virgocentric infall velocity from $v_{_{\rm LG}}$ since previous samples such as XBAC-CIZA did not include Virgo.  For a meaningful comparison between our results and those of earlier studies we follow the same approach and set $v^{\prime}_{p} = v_{p} - v_{\rm infall} = 507$ km s$^{-1}$ and $(l_{_{\rm CMB}},b_{_{\rm CMB}})$ = $(276^{\circ}, 16^{\circ})$.

The inverse of the sample's selection function, $\phi(r)$, is needed in equation (2) whenever a flux-limited catalog is employed in order to correct for the non-detection of intrinsically less luminous objects with increasing distance.  The selection function is defined as the fraction of the cluster number density that is observed above the flux limit at a given distance:
\begin{equation}
   \phi(r) = \frac{1}{{\bar{n}_{c}}} \int^{\infty}_{L_{\rm min}(r)} \Phi_{\rm X}(L)dL,
\end{equation}
where $\bar{n}_{\rm c}$ is the average cluster density, $\Phi_{\rm X}(L)$ is the X-ray cluster luminosity function and $L_{\rm min}(r) = 4\pi r^{2}S_{\rm lim}$, where $S_{\rm lim}$ is the flux limit.  We estimate $\bar{n}_{\rm c}$ by integrating the luminosity function over the entire luminosity range of the sample
\begin{equation}
                 \bar{n}_{c}  = \int^{\infty}_{L_{\rm min}} \Phi_{\rm X}(L)dL,
\end{equation}
where the lower luminosity is $L_{\rm min} = 5\times 10^{42}$ $h^{-2}$ ergs s$^{-1}$.

For the luminosity function, $\Phi_{\rm X}(L)$, we adopt a single Schechter-like function of the form
\begin{equation}
            \Phi_{\rm X}(L) = A \exp\left(-\frac{L}{L_{*}}\right)L^{-\alpha}   
\end{equation}
and fit it to our combined RBC sample.  Our best-fit values for $A, L_{*},$ and $\alpha$ are $5.47 \pm 0.2 \times 10^{-7}$ $h_{50}^{3}$ Mpc$^{-3}$, $5.35^{+1.1}_{-0.8} \times 10^{44}$ $h_{50}^{-2}$ ergs s$^{-1}$ cm$^{-2}$, and $1.71 \pm 0.07$, respectively.   Using these parameters we find $\bar{n}_{c} = 4.75 \times 10^{-5}$ $h^{3}$ Mpc$^{-3}$.  Figure 4 shows the binned X-ray luminosity functions for the REFLEX and eBCS portions of our sample, which were constructed using our recalculated X-ray fluxes and corrective weights for the eBCS.  

Finally, in addition to the corrective weights described in \S2.4, the $w_{i}$ term in equation (2) includes a component which weights clusters based on their relative mass.   We estimate the mass of each cluster in our sample through the empirical relationship $M \propto L^{3/4}_{\rm X}$ (Allen et al. 2003) which links a cluster's X-ray luminosity to its mass contained within the radius $R_{200}$, defined as the distance where the mean enclosed density is 200 times the critical density of the universe at the redshift of the cluster.  We perform our dipole analysis with and without these mass weights and present both sets of results.

\subsection{Shot Noise}

The use of discrete objects such as clusters to trace the underlying density field introduces a level of shot noise to the dipole calculation.  We estimate the amount of noise introduced by the sparseness of the RBC sample by the method of Hudson (1993):
\begin{equation}
\sigma^{2}_{\rm sn} = \left(\frac{H_{o}\beta}{4\pi \bar{n}}\right)^{2} \hspace{.05in} \sum_{i=1}^{N} \left(\frac{w_{i}\textit{\textbf{\^{r}}}_{i}}{\phi(r_{i})r_{i}^{2}}\right)^{2},
\end{equation}
such that $\sigma^{2}_{\rm sn}$ is the root-mean-square (rms) of the cumulative variance of the dipole vector sum.   Assuming that the shot-noise variance along each dipole component is equal, the mean one-dimensional error is $\sigma_{\rm 1D} = \sigma_{\rm sn} / \sqrt3$.  We find that shot-noise produces roughly $20\%$ of the RBC cluster dipole signal at 300 $h^{-1}$ Mpc (number-weighted amplitude); this is roughly a $40\%$ reduction of the shot-noise contribution in the sparser XBACs-CIZA dipole.  It should be noted that while we calculate and plot the shot-noise amplitude along with the cluster dipole, we do not subtract the shot-noise component from the dipole amplitudes shown in figures 4, 5 and 7.

\section{Results and Discussion}

The measured amplitude of the dipole anisotropy present in the composite RBC sample, in both the number- and mass-weighting schemes, is shown in Figures 5 and 6, respectively.  The dipole amplitude at any given distance is directly proportional to the peculiar velocity that is induced on the LG by the cluster distribution within that distance.  The amplitude is shown in both the LG and CMB rest frames; these should be taken as upper and lower estimates of the true dipole (Branchini \& Plionis 1996).  The direction of the dipole in both rest frames and weighting schemes is shown in Figure 7.  

\begin{figure}[t]
\epsscale{1.20}
\plotone{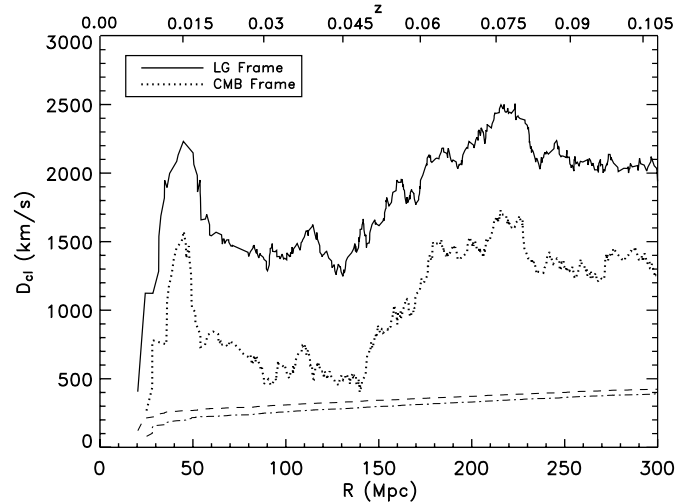}
\caption{The number-weighted RBC X-ray cluster dipole amplitude versus distance in both the LG and CMB frames (solid and dashed lines, respectively).  The dashed and dashed-dotted lines show the shot-noise amplitude in the LG and CMB frames, respectively.}
\end{figure}

\begin{figure}
\epsscale{1.20}
\plotone{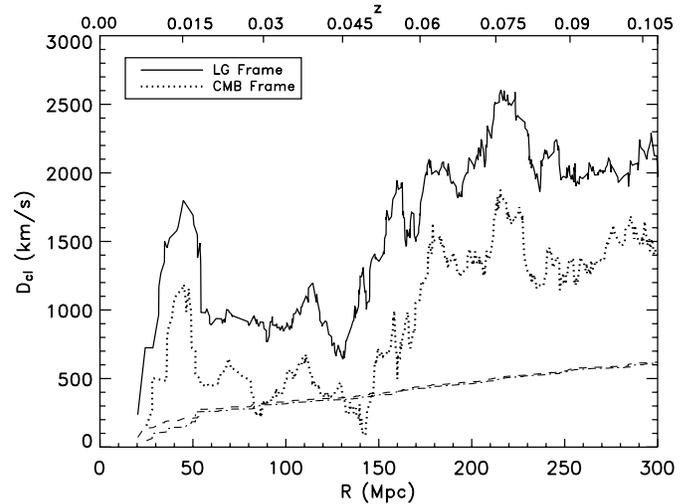}
\caption{The mass-weighted RBC X-ray cluster dipole amplitude versus distance in both the LG and CMB frames (solid and dotted lines, respectively).  The dashed and dashed-dotted lines show the shot-noise amplitude in the LG and CMB frames, respectively.}
\end{figure}

In agreement with previous cluster dipole analyses, we find that the amplitude of the anisotropy in the RBC sample is dominated by cluster concentrations at $\sim40$ and $\sim150$ $h^{-1}$ Mpc.  These distances largely match the locations of cluster overdensities in the GA and SSC regions.  Unlike the anisotropy in galaxy samples such as the PSCz and 2MRS, which show only a negligible contribution to the LG's dynamics from distances beyond 60 $h^{-1}$ Mpc, we find that the RBC dipole amplitude continues to grow until roughly $180$ $h^{-1}$ Mpc, after which it temporarily increases and then returns to its final value.  This would suggest that the RBC sample becomes isotropic with respect to the LG in the distance range of $180$ to $240$ $h^{-1}$ Mpc.  The large increase in the amplitude at $\sim150$ $h^{-1}$ Mpc implies that clusters beyond the GA induce a significant acceleration on the LG, which would seem to support studies that find evidence for a large amplitude bulk flow continuing beyond the GA.  Averaging the fraction of the number-weighted amplitude generated from beyond 60 $h^{-1}$ Mpc in both rest frames, we estimate $49\%$ of the LG's peculiar velocity originates from beyond the Hydra-Centaurus complex.  This contribution rises to $63\%$ if the clusters are weighted by their mass.  The increased large-scale contribution in the mass-weighted dipole is due to the fact that distant RBC clusters are intrinsically more luminous than their nearby counterparts (due to the flux-limited nature of the sample) and therefore have an increased weight in the mass-weighting scheme, which in turn increases the overall contribution from beyond 60 $h^{-1}$ Mpc.    Averaging the results in both rest frames and weighting schemes, we find that roughly $56\%$ of the LG's peculiar velocity is produced by clusters beyond $60$ $h^{-1}$ Mpc.

\begin{figure}[t]
\vspace*{0.15in}
\epsscale{1.20}
\plotone{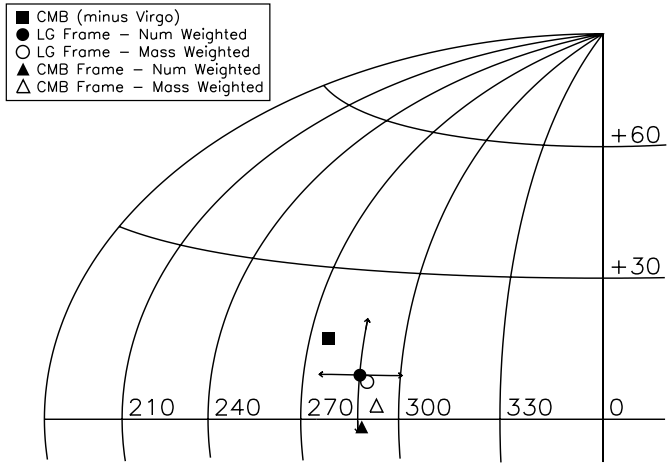}
\caption{The direction of the dipole anisotropy in Galactic coordinates.  The filled square is the direction of the CMB dipole corrected for a 170 km s$^{-1}$ Virgocentric infall.  The pointing error is computed from the directional uncertainty introduced by the shot-noise dipole.}
\end{figure}

\begin{figure}
\epsscale{1.20}
\plotone{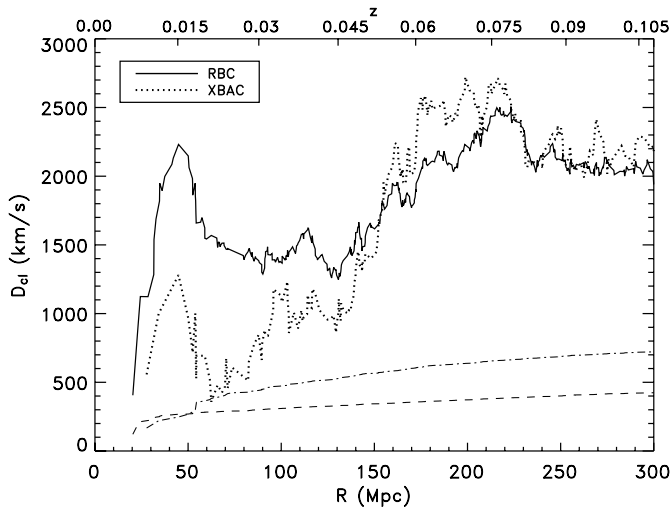}
\caption{The number-weighted RBC and XBAC X-ray cluster dipole amplitude versus distance in the LG rest frames (solid and dotted lines, respectively).  The dashed and dashed-dotted lines show the shot-noise amplitude for RBC and XBAC samples, respectively.}
\end{figure}

We also find that the misalignment between the CMB and cluster dipoles varies little throughout our study volume, implying that an anisotropy in the direction of the CMB dipole is present in the cluster distribution as early as the GA region $\sim40 h^{-1}$ Mpc away.  This agrees with previous studies which found a large-scale anisotropy in both the optically-selected cluster distribution (Plionis \& Valdarnini 1991) and infrared-selected galaxy samples (Basilakos \& Plionis 1998).  After encompassing the GA and Perseus-Pegasus supercluster (also known as Perseus-Pisces, hereafter PP), the cluster dipole is pointed within $13^{\circ}$ of the CMB dipole direction (LG-frame, number-weighting).  After 180 $h^{-1}$ Mpc the dipole in both rest-frames and weighting schemes makes its closest approach to the CMB dipole direction at roughly 220 $h^{-1}$ Mpc, with the smallest misalignment angle, $14^{\circ}$, being achieved in the number weighted, LG-frame calculation.  The misalignment angle then remains fairly constant until roughly 250 $h^{-1}$ Mpc, after which it steadily rises, presumably due to the increased shot-noise beyond that distance. 

Our results are in good agreement with the findings of Plionis \& Kolokotronis (1998) and Kocevski et al. (2004), who measured the dipole in the sparser XBAC and XBAC-CIZA samples, respectively.  The XBAC-CIZA dipole amplitude is shown with the results of this work in Figure 8.  The qualitative similarity between the RBC and XBAC-CIZA dipoles is not a surprise given that the latter sample is a subset of the former and therefore traces, to a large extent, the same superclusters that influence the LG's peculiar velocity.  The primary difference between our results and those of Kocevski et al. (2004) is a slightly reduced large-scale contribution to the number-weighted dipole amplitude:  we find 49\% of the dipole is induced from beyond $>60$ $h^{-1}$ Mpc using the RBC sample versus 62\% using XBAC-CIZA.  The reason for this difference stems from the XBAC's known incompleteness at low redshifts, where very extended clusters are systematically missed due to the optical selection of the catalog.  The XBAC sample contains 32 clusters within 100 $h^{-1}$ Mpc, whereas the RBC has 82 within the same volume.  The undersampling of XBAC clusters at low redshifts leads to an increased contribution to the dipole from greater distances, where the XBAC incompleteness is minimal.  Despite the difference in the dipole amplitude, our findings seem to confirm the general conclusions reached by Kocevski et al. (2004): a significant component of the LG's peculiar velocity is induced by overdensities beyond the GA. 

Despite the agreement with previous cluster-based dipole analyses, our findings differ from the results obtained using galaxy catalogs such as the PSCz and 2MRS.  Whereas we find a significant contribution to the dipole arising from 130 to 180 $h^{-1}$ Mpc, Rowan-Robinson et al. (2000) find only a marginal contribution to the PSCz dipole over the same distance range (see also Saunders et al. 1999 and Schmoldt et al. 1999a,b).   Likewise, Erdogdu et al. (2005) report that the dipole of the 2MRS sample largely converges by $60$ $h^{-1}$ Mpc.  We suspect the reason for this discrepancy is two-fold.  First, while the PSCz and 2MRS surveys sample the nearby galaxy population quite well, their redshift distributions reach a maximum near $z\sim0.02$ ($\sim60 h^{-1}$ Mpc) and rapidly decline afterwards (more so for the 2MRS than the PSCz, which has a longer redshift tail).  The RBC sample, on the other hand, has limited resolution nearby (simply due to the volume under consideration) but does not exhibit such a turnover until a much larger distance ($\sim240 h^{-1}$ Mpc).  Second, the IRAS catalog, from which the PSCz sample hails, is known to undersample the elliptical galaxy population and therefore the dense regions that our X-ray selected cluster sample traces best.  As we will discuss in the next section, most of our dipole signal near 150 $h^{-1}$ Mpc is produced by superclusters such as the SSC and the Horologium-Reticulum (HR) system.  In other words, the densest regions in the RBC sample produce most of our signal at large distances and it is precisely these regions that are undersampled by the PSCz.  We propose that poor sampling by galaxy samples of both distant superclusters and dense regions such as the SSC most likely explains the discordant results from the PSCz/2MRS and RBC dipole analyses.

\subsection{Supercluster Contributions}

The increased size of the RBC catalog over the XBAC-CIZA sample affords us greater resolution in tracing the large-scale structures which give rise to the LG's peculiar velocity.  Since the catalog probes the cluster distribution to fainter fluxes and is therefore not limited to extremely massive clusters which only sparsely sample the underlying density field, the RBC sample allows us to better discern which superclusters have the most dynamical impact on the LG.   To this end, we have examined the dipole profile on a cluster-by-cluster basis to investigate the relative contribution of various supercluster concentrations to the final dipole amplitude.  A diagnostic dipole profile, which includes the location of individual cluster and superclusters, is shown in Figure 9.  

As previously mentioned, the RBC dipole amplitude is dominated by cluster concentrations at $\sim40$ and $\sim150$ $h^{-1}$ Mpc.  The initial increase in the amplitude at $\sim40$ $h^{-1}$ Mpc is produced almost exclusively by the GA region, which is the first supercluster encountered in the dipole calculation (see Figure 10).  Within the RBC sample we find 13 clusters that are associated with the GA region (i.e. in the vicinity of Centaurus, Norma and CIZA J1324.7-5736); these make up $93\%$ of all the RBC clusters within $50$ $h^{-1}$ Mpc.  At this distance the GA overdensity creates an anisotropy that is pointed within $14^{\circ}$ of the CMB dipole.  We propose that this early alignment with the CMB dipole direction is the reason that many studies which have only probed the peculiar velocity field out to the GA distance manage to obtain a good alignment with the CMB anisotropy, despite overlooking more distant structures, such as the SSC.  Moving to larger distances we find the initial rise in the amplitude is counteracted between 50 and 60 $h^{-1}$ Mpc due to the effects of the PP supercluster which lies on the opposite side of the sky relative to the GA.  Between 70 to 130 $h^{-1}$ Mpc the dipole varies little in both amplitude and direction, with the largest fluctuations arising from the addition of the Lepus, Hercules and Pisces-Cetus (PC) superclusters.  The former two overdensities are responsible for the rise and fall of the amplitude between 100 and 120 $h^{-1}$ Mpc, while the latter creates the downturn between 120 and 130 $h^{-1}$ Mpc.

The next supercluster to be encompassed beyond 130 $h^{-1}$ Mpc is the SSC, which is by far the densest region in the RBC sample.  Although the increase near $150 h^{-1}$ Mpc is largely due to the SSC, we find that its origin is more complex than the signal at 40 $h^{-1}$ Mpc which was solely due to a single supercluster.  To investigate the source of the increase we identify each cluster which produces a positive jump in the amplitude between $130$ and $180$ $h^{-1}$ Mpc and determine its association to known and unknown overdensities.  We find that a total of 75 clusters contribute to the increase, of which 54 are REFLEX clusters, 12 are CIZA members and 7 belong to the BCS sample.  We can associate 17 of these clusters with the SSC region and 13 to the HR supercluster (i.e. near Abell 3128)\footnote{hereafter we shorten Abell to 'A' and refer to CIZA clusters simply by their right ascension, e.g., C1324}.  In addition to these well known regions, there exists numerous groupings and loose associations at roughly the same distance that have a significant effect on the amplitude.  The most notable among these is a previously unknown association of clusters around A3667.  While A3667 is a well known merging cluster (Knopp et al. 1996, Markevitch et al. 1999), we can find no reference of its participation in an overdensity such as a supercluster.  This is not surprising since A3667 has only two Abell cluster moderately nearby, A3651 and A3716.  Kocevski et al. (2004) noticed A3667 since its sizable mass and favorable location provided a boost to the XBAC-CIZA dipole, but the cluster was not associated with a group since it has only one companion in the XBAC catalog (i.e. A3716).   With the reduced flux limit of the RBC sample, we can now resolve that A3667 is actually the core member of a loose association of 7 clusters.  Another notable group is a set of 6 clusters centered on A3391, which lies between HR and the ZOA.  The group contains A3391, A3395, and A3380 and may be an extension of the larger HR region located 50 $h^{-1}$ Mpc away (see Figure 10).  Also of interest are numerous CIZA clusters which cross the ZOA between $130$ and $180$ $h^{-1}$ Mpc.  One notable structure is a string of 6 CIZA clusters that traverse the ZOA at the same distance as the SSC and the A3667 association.  One these clusters, C1410, is near enough to the SSC to be considered a member of the supercluster; the smooth transition from SSC members to CIZA clusters in this region suggests this string of clusters may trace an extension of the filament network in which the SSC is embedded into the ZOA.  This association also includes the well-known Triangulum Australis cluster (C1638) and the CIZA cluster C1652, which was noted by Kocevski et al. (2004); we will refer to this region as the C1410 Filament hereafter.  Another set of interesting CIZA members are 3 clusters near the outer portions of the A3391 group which may  hint at an extension of this association behind the plane of the Galaxy; the groups central cluster is C0821.  The locations of these superclusters is displayed in supergalactic coordinates in Figure 10.

We next calculate the contribution of each overdensity to the increase at $150 h^{-1}$ Mpc; our results are listed in Table 1.  We find the SSC is the greatest single contributor to the increase, with its clusters producing $30.4\%$ of the signal.  The C1410 Filament is second, contributing $12.5\%$ to the overall increase\footnote{This total does not include the effects of C1410, which is included in the SSC contribution}.  The C1821 Extension produces $9.3\%$ of the jump near $150 h^{-1}$ Mpc, while the A3391 association induces the next largest fraction at $8.6\%$.  The final two regions, the HR and A3667 associations, produce $8.3\%$ and $5.8\%$, respectively.  It should be noted that if one considers the A3391 group an extension of HR region, the combined system would be second only to the SSC in its effect of the dipole amplitude, contributing a total of $16.9\%$.  Although the HR is further south than the SSC and on the opposite side of the ZOA (see Figure 10), the location of the supercluster is near enough to the dipole direction at 150 $h^{-1}$ Mpc to provide a sizable contribution to the increase.  In total these cluster associations account for $71.9\%$ of the large-scale contribution to the cluster dipole.  The remaining $28.1\%$ is produced by isolated clusters in the general direction of the resultant dipole pointing near $150 h^{-1}$ Mpc. 

\begin{center}
\begin{deluxetable}{lrr}
\tablenum{1}
\tablewidth{2 in}

\tablecaption{Contributions to the dipole amplitude between $130$ and $180$ $h^{-1}$ Mpc}

\tablecolumns{3}
\tablehead{  \colhead{Supercluster} & \colhead{\# Clusters} & \colhead{\% Contrib.}   }
\startdata
Shapley      &  17  &  30.4  \nl
C1410 Filam. &   5  &  12.5  \nl
C1821 Ext.   &   3  &   9.3  \nl
A3391 Assoc. &   6  &   8.6  \nl
Hor-Ret.     &  13  &   8.3  \nl
A3667 Assoc. &   7  &   5.8  \nl\nl
\hline \nl
All REFLEX   &  54  &  65.4  \nl
All CIZA     &  12  &  29.6  \nl
All eBCS     &   7  &   5.0  \nl
\enddata
\end{deluxetable}
\end{center}

\vspace*{-0.3in}

One final note regarding the origin of the increase near $150$ $h^{-1}$: while it is evident that the presence of numerous superclusters and overdensities near this distance and in the direction of the LG's peculiar velocity play a large role in creating the increase in the dipole amplitude, of equal importance is the absence of clusters in the opposite direction.  Examining the redshift distribution of the eBCS and REFLEX samples, it is readily apparent that compared to REFLEX, the eBCS is underdense between $130$ and $180$ $h^{-1}$ Mpc.  We find that the region of sky within $90^{\circ}$ of the dipole pointing in this shell is 2.7 times as dense as the opposite part of the sky.  There is no reason to believe this density variation is artificial, indicating that the anisotropy in the cluster distribution near $150$ $h^{-1}$ Mpc has as much to do with the presence of large overdensities such as the SSC and the HR concentrations in the south as it does with the lack of superclusters in the north.  The fact that galaxy samples do not  detect a significant dipole signal at this distance may be in part because they do not detect as large a contrast between the density field in the northern and southern hemispheres owing to their sparse sampling of the galaxy population at these distances.

\subsection{The $\beta$ Parameter}

\begin{figure*}
\vspace*{0mm} \mbox{}\\
\hspace*{0.25in}
\epsfxsize=6in 
\epsffile{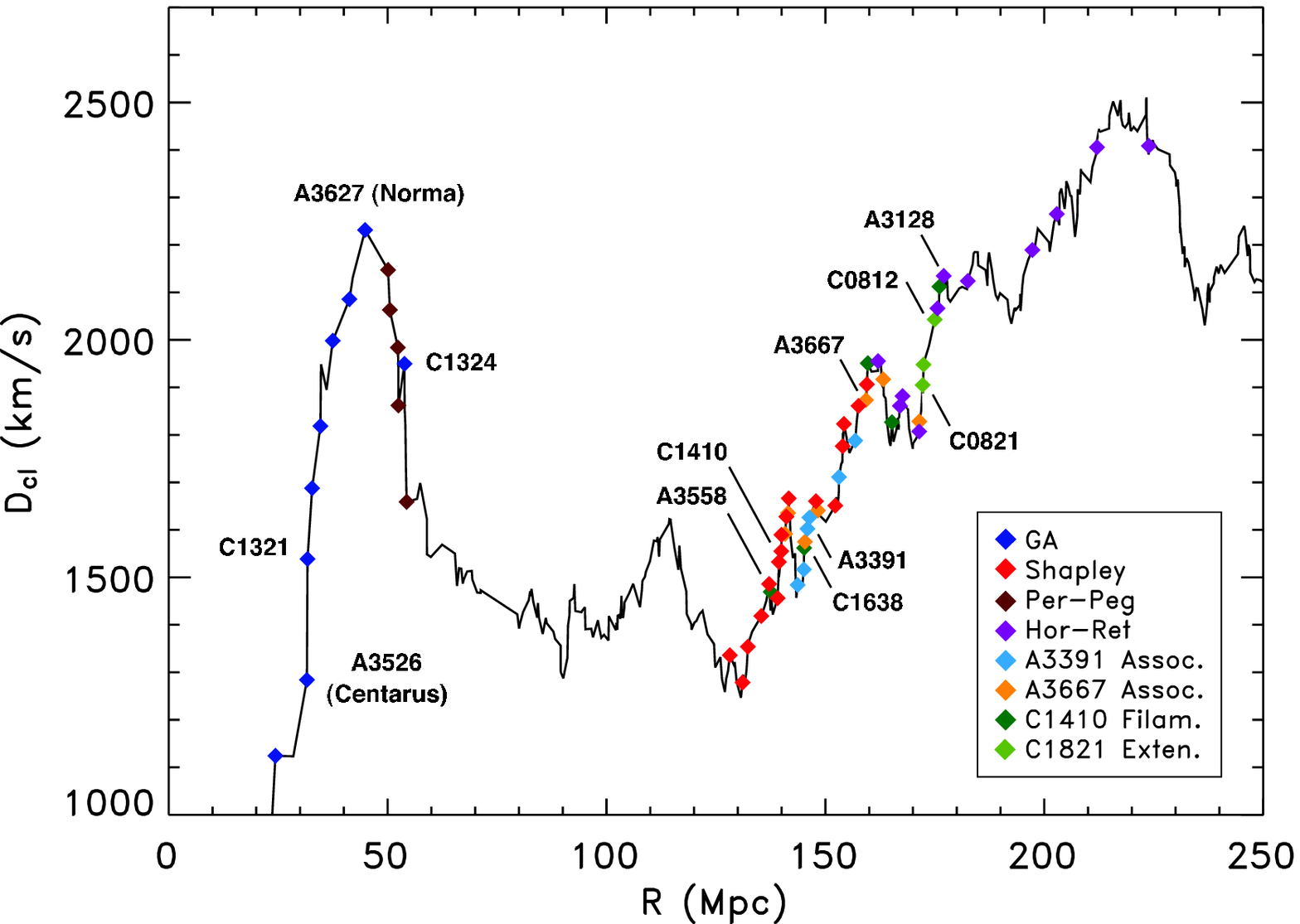}\\
\baselineskip0.6mm
Fig.~8.--- Schematic dipole profile; see text for details. Cluster associations are grouped by color to highlight their impact on the overall dipole amplitude.  Abell and CIZA clusters begin with the letters 'A' and 'C', respectively.  Acronyms are GA: Great Attractor, Hor-Ret: Horologium-Reticulum, Per-Peg: Perseus-Pegasus.  \\
\end{figure*}

\begin{figure*}
\vspace*{0mm} \mbox{}\\
\hspace*{0.4in}
\epsfxsize=6in 
\epsffile{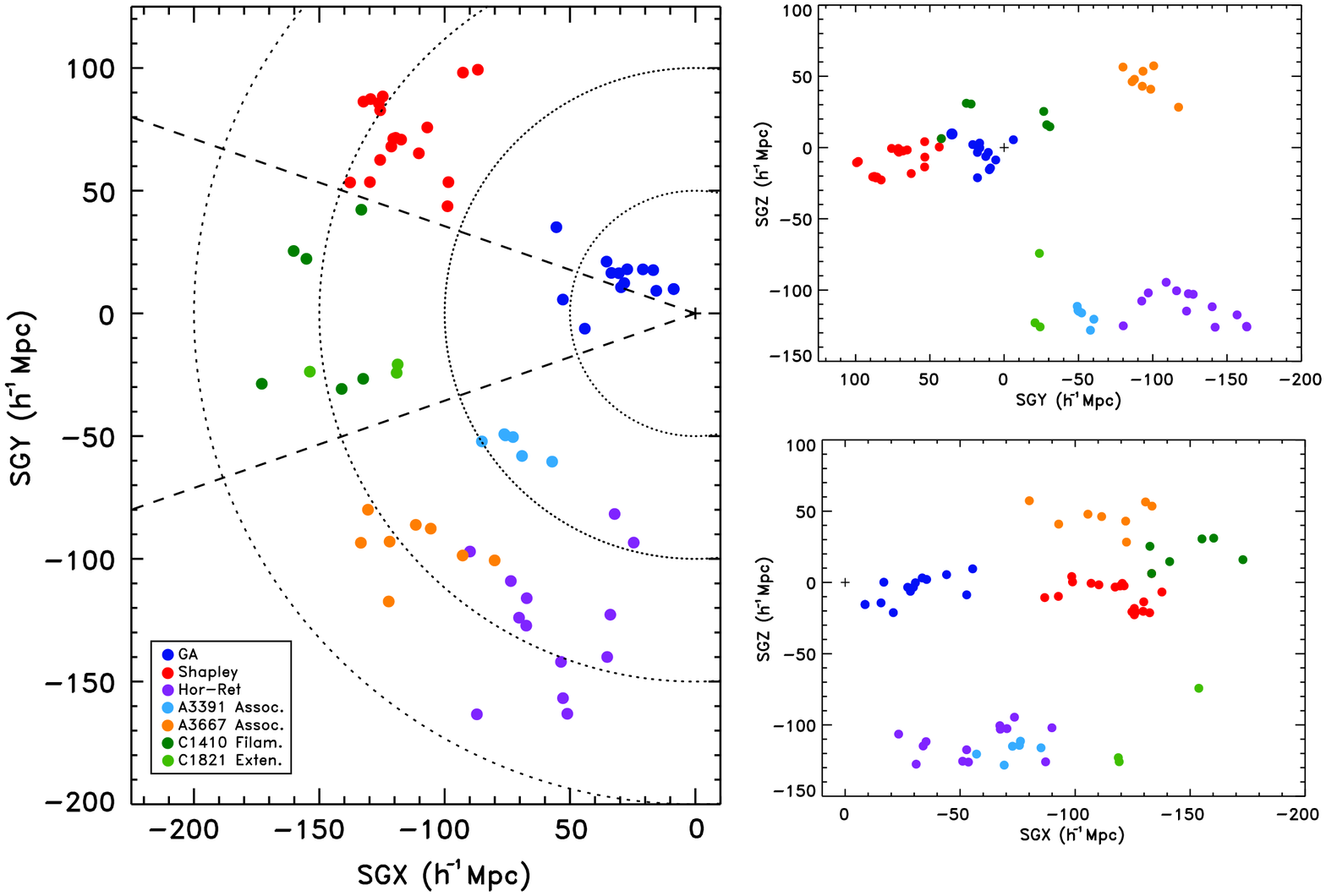}\\
\baselineskip0.6mm
Fig.~9.--- The location of cluster associations highlighted in Figure 9 as projected into the supergalactic coordinate system.  Colors and acronyms are the same as in Figure 9.  The dashed line represents the boundary of the traditional ZOA ($|b|<20^{\circ}$), while the dotted lines demark 50 Mpc steps.  With the exception of the HR and A3391 associations, the structures which are largely responsible for inducing the LG's motion lie on a relatively thin SGX-SGZ plane. \\
\end{figure*}

By comparing the peculiar velocity of the LG as inferred from the CMB dipole, $\textit{v}_{p}$, to that predicted by the X-ray cluster distribution, $\textit{D}_{\rm cl}$ we can use Equation 2 to determine the biasing parameter, $\beta = \Omega^{0.6}_{0}/b$, which relates the RBC clusters to the underlying mass distribution which they trace.  Although the Virgo cluster is included in the RBC sample, authors have traditionally removed the infall velocity of the LG toward Virgo since previous samples such as XBAC-CIZA did not include the Local Supercluster.  For a meaningful comparison between our results and those of Plionis \& Kolokotronis (1998) and Kocevski et al. (2004) we follow the same approach.

To obtain $\textit{D}_{\rm cl}$, we calculate the median of the dipole amplitude between 240 and 300 $h^{-1}$ Mpc, where we assume isotropy has been reached and the amplitude has arrived at its final value.  Without Virgo, the LG and CMB reference frame, number-weighted amplitudes are $2070 \pm 53$  and $1321 \pm 62$ km s$^{-1}$, respectively, where the errors listed are from the variation of the amplitude over the given range.  To remove the LG's Virgocentric infall velocity from its peculiar motion we use
\begin{equation}
	 \textit{v}^{\prime}_{p} = v_{p} - v_{\rm inf}cos(\delta\theta)
\end{equation}
where $v_{p}$ is 627 km s$^{-1}$, $\delta\theta$ is the angle between the CMB dipole and Virgo directions and is roughly $45^{\circ}$ and we set $v_{\rm inf}$, the infall velocity, to the literature average value of 170 km s$^{-1}$, as used by Plionis and Kolokotronis (1998) and Kocevski et al. (2004).  With these values we obtain a corrected peculiar velocity of $v^{\prime}_{p} = 507$ km s$^{-1}$.  Dividing this by the dipole amplitudes we obtain the following upper and lower estimates on the $\beta$ parameter:
\begin{eqnarray}
		\beta_{\rm LG}  = 0.24\pm 0.01 \\
  		\beta_{\rm CMB} = 0.38\pm 0.02
\end{eqnarray}
where the errors are again from the variation of the amplitude over the distance of 240 and 300 $h^{-1}$ Mpc.  Performing the same analysis on the mass-weighted amplitude we find $\beta_{\rm LG} = 0.25\pm 0.01$ and $\beta_{\rm CMB} = 0.37\pm 0.03$.

\section{Conclusions}
We have combined the REFLEX catalog in the southern hemisphere, the eBCS sample in the north, and the CIZA survey in the Galactic plane to produce the largest, all-sky, truly X-ray selected cluster sample compiled to date and used it to investigate the origin of the LG's peculiar velocity.  The X-ray selected nature of the RBC catalog largely does away with projection effects, the poor sampling of nearby clusters, and the incompleteness in the plane of the Galaxy introduced as a result of the optical selection methods used to construct previous cluster catalogs.  In addition, the RBC's relatively low X-ray flux limit allows us to better sample the underlying density field, which in turn affords us a greater resolution in tracing the structures which give rise to the LG's motion.  

From the dipole anisotropy present in the cluster distribution we determine that $44\%$ of the LG's peculiar velocity is due to infall into the GA region, while $56\%$ is induced by more distant overdensities between $130$ and $180$ $h^{-1}$ Mpc away.  Of the large-scale contribution, we find that the SSC has the largest dynamical impact, being responsible for $30.4\%$ of the increase in the dipole amplitude beyond $130$ $h^{-1}$ Mpc.  Despite the significance of the GA and SSC regions, our findings are not consistent with a simple two-attractor model for generating the LG's motion.  There exists numerous groupings and loose associations of clusters at roughly the same distance as the SSC that have a significant effect on the dipole amplitude.  These include the well-known HR region, as well as the newly noted A3667 and A3391 associations and the C1410 Filament which may trace an extension of the SSC complex into the ZOA.  In addition, we find that the region of sky within $90^{\circ}$ of the dipole pointing between $130$ and $180$ $h^{-1}$ Mpc is 2.7 times as dense as the opposite part of the sky.  We suggest that the anisotropy in the cluster distribution near $150$ $h^{-1}$ Mpc has as much to do with the presence of large overdensities such as the SSC and the HR concentrations in the south as it does with the lack of superclusters in the north.

\acknowledgments
We thank Brent Tully and Chris Mullis for many valuable discussions and contributions, as well as Manolis Plionis for his useful suggestions.  DK is supported by the NASA Graduate Student Research Program.

\end{document}